\documentclass[twocolumn,showpacs,amsmath,amssymb]{revtex4}

\usepackage{graphicx}
\usepackage{dcolumn}
\usepackage{bm}

\renewcommand{\vec}[1]{\mbox{\boldmath $#1$}}

\begin{document}

\preprint{}

\title{Deformation of $\Lambda$ hypernuclei}

\author{Myaing Thi Win}
\affiliation{
Department of Physics, Tohoku University,
Sendai 980-8578, Japan}

\author{K. Hagino}
\affiliation{
Department of Physics, Tohoku University,
Sendai 980-8578, Japan}

\date{\today}

\begin{abstract}

We study the deformation property of $\Lambda$ hypernuclei 
using the relativistic mean field (RMF) method. 
We find that 
$^{29}_{~\Lambda}$Si and $^{13}_{~\Lambda}$C hypernuclei have spherical 
shape as a consequence of the additional $\Lambda$ particle, 
whereas the corresponding core nuclei, 
$^{28}$Si and $^{12}$C, are oblately deformed. 
Most of other hypernuclei have a similar deformation 
parameter to the core nucleus, in accordance with the previous 
study with the non-relativistic Skyrme-Hartree-Fock method. 
We discuss the sensitivity of our results to the choice of 
pairing interaction and to the parameter set of the RMF Lagrangian. 
\end{abstract}

\maketitle

\section{Introduction}

It has been well known that many open-shell nuclei are deformed in the 
ground state. The nuclear deformation generates the collective 
rotational motion, which is characterized by 
a pronounced rotational spectrum as well as 
strongly enhanced 
quadrupole transition probabilities. 
Theoretically, a standard way to discuss nuclear deformation is 
a self-consistent mean-field theory\cite{BHR03}. 
By allowing the rotational symmetry to be broken in the mean-field potential, 
the mean-field theory provides an intuitive and transparent view of the 
nuclear deformation. 
See {\it e.g.,} Ref. \cite{SDNPD03} 
for a recent systematic study of nuclear deformation 
based on the Skyrme-Hartree-Fock-Bogoliubov method. 
The state-of-the-art mean-field approach also takes into account the effects 
beyond the mean-field approximation, such as the angular momentum projection 
and the configuration mixing \cite{BH08}. 

The self-consistent mean-field method has been extensively applied also to 
hypernuclei
\cite{R81,YBZ88,YB90,LY97,L98,CLS00,VPRS01,ZSSWZ07,MZ89,RSM90,GVH93,MJ94,ST94,VPLR98,LMZZ03,SYT06} (see Ref. \cite{HT06} for a recent experimental 
review on $\Lambda$ hypernuclei). 
These calculations have successfully reproduced the mass number 
dependence of $\Lambda$ binding energy, from a 
light nucleus $^{12}_{~\Lambda}$C 
to a heavy nucleus $^{208}_{~~\Lambda}$Pb. 
We notice that most of these calculations have assumed spherical symmetry. 
Only recently, deformed calculations have been carried out in a broad 
mass region using the 
non-relativistic Skyrme Hartree-Fock method \cite{ZSSWZ07}. 
The authors of Ref. \cite{ZSSWZ07} have reported that the hypernuclei 
which they studied have a similar deformation parameter 
to the corresponding core nuclei with the same sign. 

The aim of this paper is to study the deformation property of 
$\Lambda$ hypernuclei using the relativistic mean field (RMF) method,  
as an alternative choice of effective $NN$ and $N\Lambda$ interactions. 
The RMF method has been as successful as the Skyrme-Hartree-Fock 
method in describing stable nuclei as well as 
nuclei far from the stability line\cite{RGL90,R96}. 
Vretenar {\it et al.} have argued \cite{VPLR98} 
that the change in the nucleon spin-orbit 
potential due to the presence of $\Lambda$ particle is much 
more emphasized in the RMF approach as compared to the non-relativistic 
approach, since the spin-orbit potential in the former approach is actually 
given as a sum of scalar and vector potentials. 
That is, even if the change in the mean-field potential (given as 
a difference of scalar and vector potentials) is small,  
the change in the spin-orbit potential may not necessarily be small. 
Therefore, a slightly different conclusion from that with 
the non-relativistic approach may result concerning 
the structure of hypernuclei. 
In fact, we will demonstrate below that the shape of 
$^{12}$C and $^{28}$Si nuclei are drastically changed when a $\Lambda$ 
particle is added to them. 

The paper is organised as follows. In Sec. II, we briefly summarize the 
RMF approach for $\Lambda$ hypernuclei. In Sec. III, we apply the RMF 
method to Ne and Si isotopes, and discuss the influence of $\Lambda$ 
particle on the deformation of the hypernuclei. We also discuss the 
deformation of $^{12}$C and $^{13}_{~\Lambda}$C nuclei. We summarize 
the paper in Sec. IV. 

\section{RMF for $\Lambda$ hypernuclei}

In the RMF approach, nucleons and a $\Lambda$ particle are treated as 
structureless Dirac particles, interacting through the exchange of virtual 
mesons, that is, the isoscalar scalar $\sigma$ meson, 
the isoscalar vector $\omega$ meson, and the isovector vector $\rho$ meson. 
The photon field is also taken into account to describe the Coulomb 
interaction between protons. 
The effective Lagrangian for $\Lambda$ hypernuclei may be given 
as \cite{MZ89,RSM90,GVH93,MJ94,ST94,VPLR98,LMZZ03,SYT06} 
\begin{equation}
{\cal L}={\cal L}_N+\bar{\psi}_\Lambda\left[
\gamma_\mu\left(i\partial^\mu-g_{\omega\Lambda}\omega^\mu\right)
-m_\Lambda-g_{\sigma\Lambda}\sigma\right]\psi_\Lambda,
\label{RMF}
\end{equation}
where $\psi_\Lambda$ and 
$m_\Lambda$ are 
the Dirac spinor and the mass for the $\Lambda$ particle, respectively. 
Notice that the $\Lambda$ particle couples only to the 
$\sigma$ and $\omega$ mesons, as it is neutral and isoscalar. 
Those coupling constants are denoted as 
$g_{\sigma\Lambda}$ and $g_{\omega\Lambda}$, respectively. 
For simplicity, we neglect the tensor $\Lambda$-$\omega$ interaction. 
This is justified since we consider only the ground state 
configuration, in which the $\Lambda$ particle occupies 
the lowest $K^\pi=1/2^+$ single-particle state\cite{VPLR98,ST94}, 
$K$ being the 
projection of the single-particle angular momentum onto the symmetry axis. 
${\cal L}_N$ in Eq. (\ref{RMF}) is the standard RMF Lagrangian for the 
nucleons. See {\it e.g.,} Refs. \cite{VPLR98,RGL90,R96} for its explicit form. 

We solve the RMF Lagrangian (\ref{RMF}) in the mean field approximation. 
The variational principle leads to the Dirac equation for the $\Lambda$ 
particle, 
\begin{equation}
\left[-i\vec{\alpha}\cdot\vec{\nabla}
+\beta\,
(m_\Lambda+g_{\sigma\Lambda}\sigma(\vec{r}))+g_{\omega\Lambda}\omega^0(\vec{r})
\right]\psi_\Lambda=\epsilon_\Lambda\psi_\Lambda,
\end{equation} 
where $\epsilon_\Lambda$ is the single-particle energy for the $\Lambda$ 
particle state, 
and the Klein-Gordon equation for the mesons, 
\begin{eqnarray}
[-\vec{\nabla}^2+m_\sigma^2]\sigma(\vec{r})&=&-g_\sigma\rho_s(\vec{r})
-g_2\,\sigma(\vec{r})^2-g_3\,\sigma(\vec{r})^3 \nonumber \\
&&-g_{\sigma\Lambda}\psi_\Lambda^\dagger(\vec{r})\gamma^0\psi_\Lambda(\vec{r}), 
\end{eqnarray}
\begin{equation}
[-\vec{\nabla}^2+m_\omega^2]\omega^0(\vec{r})=g_\omega\rho_v(\vec{r})
+g_{\omega\Lambda}\psi_\Lambda^\dagger(\vec{r})\psi_\Lambda(\vec{r}).
\end{equation}
To derive these equations, we have used the time-reversal symmetry and 
retained only the time-like component of $\omega^\mu$ \cite{RGL90}. 
$\rho_s$ and $\rho_v$ are the scalar and vector densities for the nucleons, 
which are constructed with the spinor for the nucleons using 
the so called no-sea approximation, {\it i.e.,} neglecting the contribution 
from the antiparticles. 
$g_\sigma$ and $g_\omega$ are the coupling constants of the nucleons to 
the sigma and the omega mesons, respectively, and $g_2$ and $g_3$ are 
the coefficients in the non-linear sigma terms in ${\cal L}_N$. 

We solve these equations, together with the Dirac equation for the 
nucleons and the Klein-Gordon equations for the $\rho$ meson and the 
photon field, iteratively until the self-consistency condition is 
achieved. 
For this purpose, we 
modify the computer code {\tt RMFAXIAL} \cite{RGL97} 
to include the $\Lambda$ particle. In this code, 
the RMF equations for the nucleons are solved 
with the harmonic oscillator expansion
method\cite{RGL90}, assuming the axial symmetry. 
The pairing correlation among the nucleons is also  
taken into account in the BCS approximation. 

With the self-consistent solution of the RMF equations, we compute the 
intrinsic quadrupole moment of the hypernucleus,
\begin{equation}
Q=\sqrt{\frac{16\pi}{5}}\,\int d\vec{r}\,[\rho_v(\vec{r})+
\psi_\Lambda^\dagger(\vec{r})\psi_\Lambda(\vec{r})]\,r^2Y_{20}(\hat{\vec{r}}).
\end{equation}
The quadrupole deformation parameter $\beta_2$ is then estimated with 
the intrinsic quadrupole moment as \cite{RGL90,RS80,HLY06},
\begin{equation}
Q =\sqrt{\frac{16\pi}{5}}\,\frac{3}{4\pi}\,(A_c+1)R_0^2\,\beta_2,
\end{equation}
where $A_c=A-1$ is the mass number of the core nucleus for the hypernucleus. 
We use $R_0=1.2A_c^{1/3}$ fm for the radius of the hypernucleus. 

\section{Quadrupole Deformation of $\Lambda$ hypernuclei}

We now numerically solve the RMF equations and discuss the quadrupole 
deformation parameter of $\Lambda$ hypernuclei. 
For this purpose,  
we use the NL3 parameter set\cite{nl3} for the 
RMF Lagrangian for the nucleons, ${\cal L}_N$. 
For the $\Lambda$-meson coupling constants, we follow 
Refs. \cite{MJ94,VPLR98} and take $g_{\omega\Lambda}=\frac{2}{3}g_{\omega}$ 
and $g_{\sigma\Lambda}=0.621g_{\sigma}$. The value for 
$g_{\omega\Lambda}$ was determined from the naive quark model \cite{ST94}, 
while the value for $g_{\sigma\Lambda}$ was slightly fine-tuned in order 
to reproduce the $\Lambda$ binding energy of $^{17}_{~\Lambda}$O \cite{MJ94}. 
For the pairing correlation among the nucleons, 
we employ 
the constant gap approximation with the pairing gap
given in Ref. \cite{MN92}, that is, 
$\Delta_n=4.8/N^{1/3}$ and $\Delta_p=4.8/Z^{1/3}$ MeV for the neutron 
and the proton pairing gaps, respectively. 
(It has been known that these pairing gaps underestimate the deformation 
parameter of $^{20}$Ne nucleus when it is calculated with the NL3 parameter 
set\cite{BG03,LRR99}. We therefore arbitrarily 
switch off the pairing interaction when we calculate the 
$^{20}$Ne and $^{21}_{~\Lambda}$Ne nuclei.)

\begin{figure}[htb]
\begin{center}\leavevmode
\includegraphics[width=0.91\linewidth, clip]{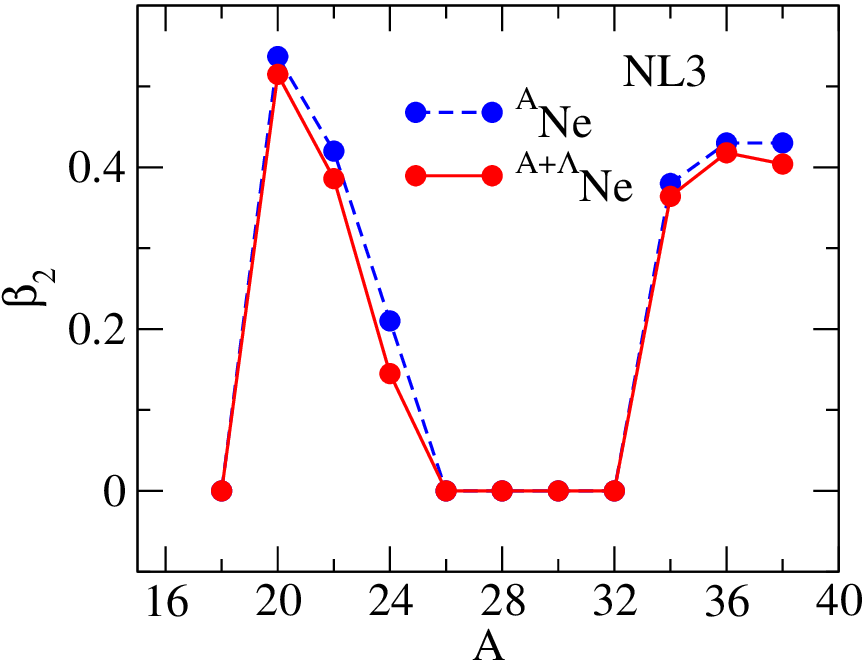}
\caption{(Color online) Quadrupole deformation parameter for Ne isotopes 
obtained with the RMF method with the NL3 parameter set. 
The dashed line is the deformation parameter for the core nucleus, while 
the solid line is for the corresponding hypernucleus. 
}
\end{center}
\end{figure}

\begin{figure}[htb]
\begin{center}\leavevmode
\includegraphics[width=0.91\linewidth, clip]{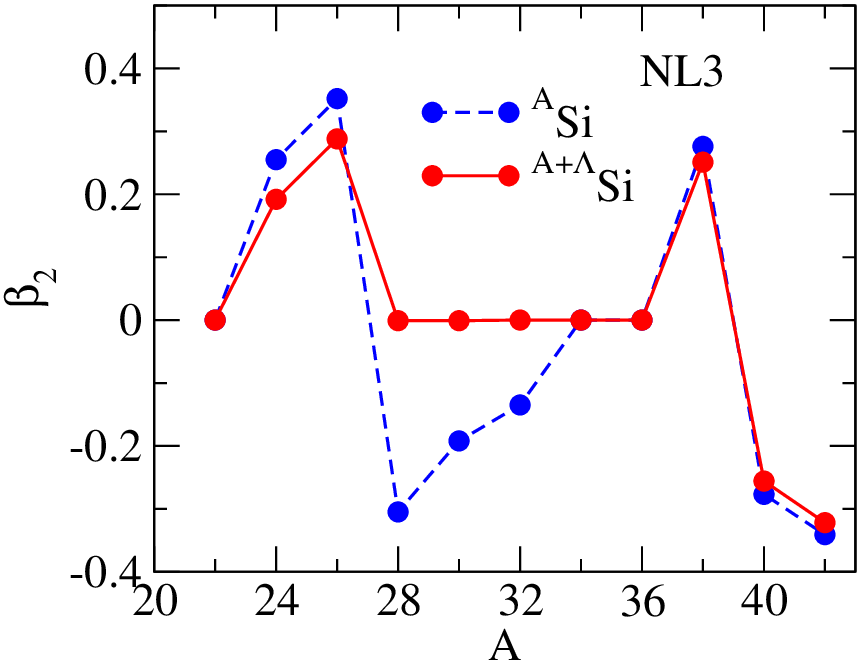}
\caption{(Color online) Same as Fig.1, but for Si isotopes. 
}
\end{center}
\end{figure}

Figures 1 and 2 show the deformation parameter for the ground state 
of Ne and Si isotopes, respectively. The dashed line is the deformation 
parameter for the even-even core nuclei, while the solid line is for 
the corresponding hypernuclei. 
For the Ne isotopes, the deformation parameter is always 
similar between the 
core nucleus and the corresponding hypernucleus, although the deformation 
parameter for the hypernucleus is slightly smaller than that for the 
core nucleus. This 
is consistent with the previous results 
with the non-relativistic Skyrme-Hartree-Fock method \cite{ZSSWZ07}. 
On the other hand, for the Si isotopes, the deformation parameter for the 
$^{28,30,32}$Si nuclei is drastically changed when a $\Lambda$ particle is 
added, although the change for the other Si isotopes is small. 
That is, the $^{28,30,32}$Si nuclei have oblate shape in the ground state. 
When a $\Lambda$ particle is added to these nuclei, remarkably they 
turn to be 
spherical. 

\begin{figure}[htb]
\begin{center}\leavevmode
\includegraphics[width=0.91\linewidth, clip]{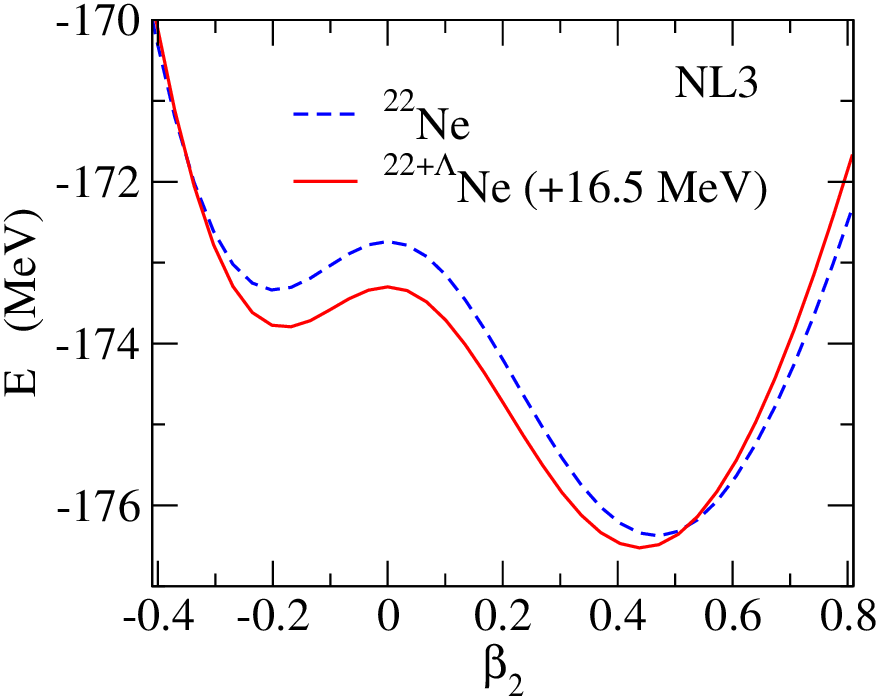}
\caption{(Color online) 
The potential energy surface for the $^{22}$Ne (the dashed line) 
and $^{22+\Lambda}$Ne (the solid line) nuclei obtained with the 
constrained RMF method with the NL3 parameter set. 
The energy surface for $^{22+\Lambda}$Ne is shifted by a constant 
amount as indicated in the inset. 
}
\end{center}
\end{figure}

\begin{figure}[htb]
\begin{center}\leavevmode
\includegraphics[width=0.91\linewidth, clip]{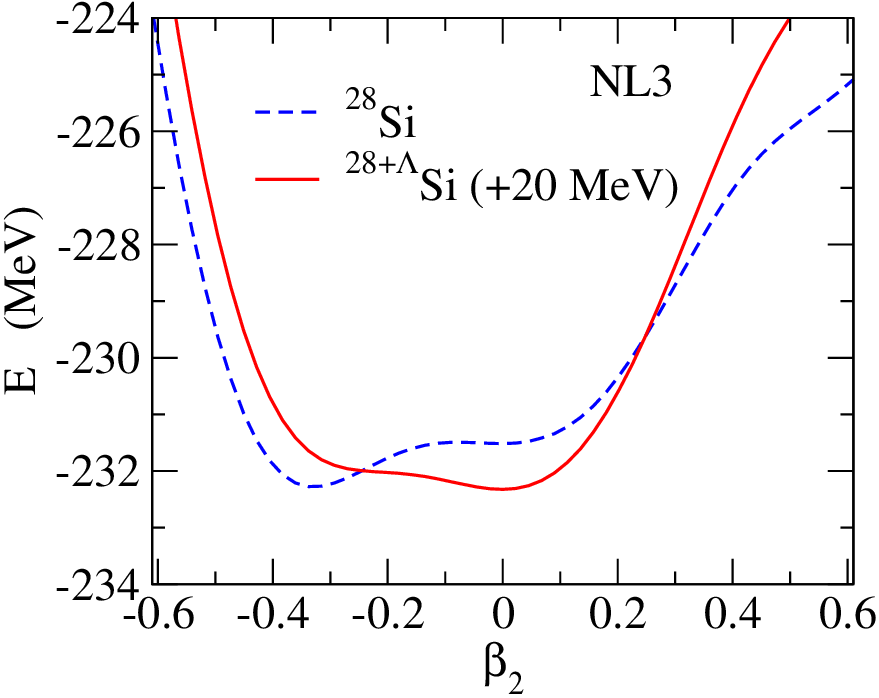}
\caption{(Color online) 
Same as Fig. 3, but for the 
$^{28}$Si and $^{28+\Lambda}$Si nuclei. 
}
\end{center}
\end{figure}

The potential energy surfaces for the $^{22,22+\Lambda}$Ne and 
$^{28,28+\Lambda}$Si nuclei are shown in Figs. 3 and 4, respectively. 
These are obtained with the constrained RMF method with quadrupole 
constraint \cite{RS80,FQKV73}. The meaning of each line is the same as 
in Figs. 1 and 2. In order to facilitate the comparison, we shift the 
energy surface for the hypernuclei by a constant amount as indicated in 
the inset of the figures. 
For the $^{22}$Ne nucleus, the prolate minimum in the energy surface is 
relatively deep (the energy difference between the oblate and the prolate 
minima is 3.04 MeV, and that between the spherical and prolate configurations 
is 3.63 MeV), and it is affected little by the addition of the 
$\Lambda$ particle. 
On the other hand, the energy surface for the $^{28}$Si nucleus shows a 
relatively shallow oblate minimum, 
with a shoulder at the spherical configuration. 
The energy difference between the oblate and the spherical configurations 
is 0.754 MeV, and could be easily inverted when a $\Lambda$ particle is 
added. 

\begin{figure}[htb]
\begin{center}\leavevmode
\includegraphics[width=0.91\linewidth, clip]{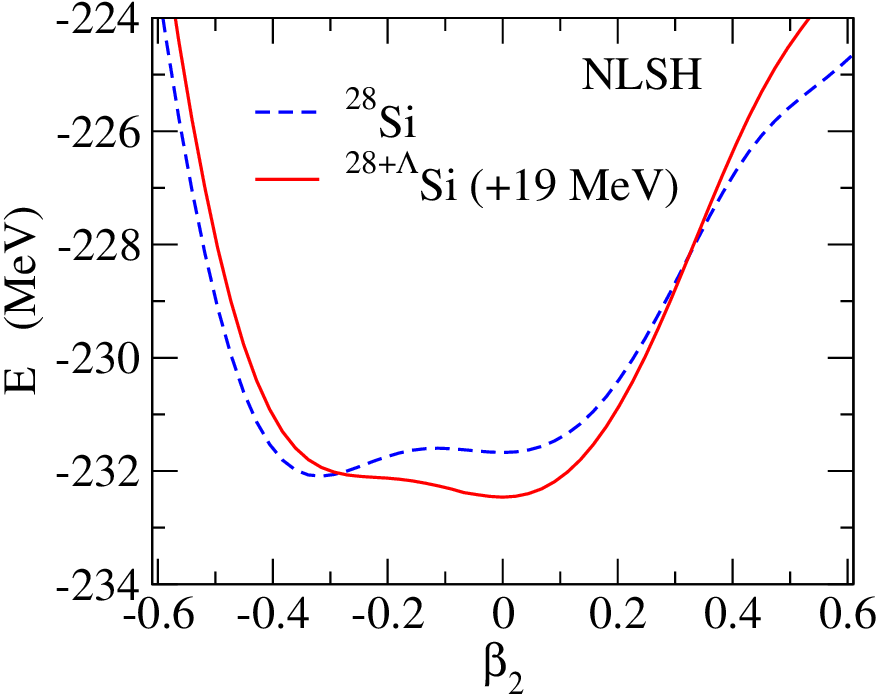}
\caption{(Color online) 
Same as Fig. 4, but 
obtained with the NLSH parameter set. 
}
\end{center}
\end{figure}

In order to check the parameter set dependence of the results, we repeat 
the same calculation with the NLSH parameter set \cite{SNR93}. 
The potential energy surface for the 
$^{28,28+\Lambda}$Si nuclei obtained with the NLSH set is shown in 
Fig. 5. One sees that the potential energy surface is qualitatively 
almost the same between the NL3 and NLSH parameter sets, although the 
$\Lambda$ binding energy is slightly different. Namely, 
the oblate $^{28}$Si nucleus becomes spherical in the presence 
of $\Lambda$ particle, again with the NLSH parameter set. 
We also check the dependence of the results on the treatment of 
pairing correlation. For this purpose, we perform the calculations i) 
without taking into account the paring correlation and ii) 
with the constant force approach 
for the strength of the seniority pairing force. 
For the latter approach, we determine
$G_p$ and $G_n$ so that they lead to 
$\Delta_n=4.8/N^{1/3}$ and $\Delta_p=4.8/Z^{1/3}$ MeV for the ground 
state of each nucleus. 
(For instance, 
$G_p=17.38/A$ and $G_n=15.97/A$ MeV for the NL3 calculation of 
$^{28}$Si nucleus.) 
We confirm that our conclusion remains the same 
for both the treatments of the pairing correlation, due to the fact 
that $N$ or $Z$=14 is an oblate magic number\cite{RNS78}. 
We therefore conclude 
that the $\Lambda$ particle significantly 
changes the deformation of $^{28}$Si nucleus, at least for the two 
parameter 
sets of the RMF Lagrangian and 
irrespective of the treatment of pairing correlations. 

\begin{table*}[hbt]
\caption{
Comparison of deformation parameter for the 
$^{28,30,32}$Si, $^{28,30,32+\Lambda}$Si, 
and $^{12,12+\Lambda}$C nuclei obtained with the NL3 and NLSH parameter 
sets of RMF Lagrangian. 
The pairing correlation is taken into account either in the constant 
$\Delta$ ({\it i.e.,} constant pairing gap) 
or in the constant $G$ ({\it i.e.,} constant pairing force) 
approximations for a seniority pairing interaction. 
The results without the pairing correlation are also shown. 
The calculation with the NL3 set 
did not converge for $^{12,12+\Lambda}$C and 
the results are not shown in the table. 
}
\begin{center}
\begin{tabular}{c|ccc|ccc}
\hline
\hline
     &      &    NL3   &    &   &  NLSH  &   \\ 
\hline
nucleus & const.-$\Delta$ &  const.-$G$ & no-pairing & 
const.-$\Delta$ &  const.-$G$ & no-pairing \\
\hline
$^{28}$Si & $-$0.31 & $-$0.29 & $-$0.33 & $-$0.29 & $-$0.25 & $-$0.32 \\
$^{28+\Lambda}$Si & 0.00 & 0.01 & 0.00 & 0.00 & 0.00 & 0.00 \\
\hline
$^{30}$Si & $-$0.19 & $-$0.19  & 0.15 & $-$0.19 & $-$0.19 & 0.19 \\
$^{30+\Lambda}$Si & 0.00 & 0.00  & 0.00 & $-$0.06 & 0.06 & 0.18 \\
\hline
$^{32}$Si & $-$0.14 & $-$0.14  & $-$0.20 & $-$0.15 & $-$0.15 & $-$0.20 \\
$^{32+\Lambda}$Si & 0.00 & $-$0.00  & $-$0.18 & $-$0.11 & 0.12 & $-$0.18 \\
\hline
$^{12}$C &  &   &  & $-$0.25 & $-$0.25 & $-$0.286 \\
$^{12+\Lambda}$C &  &  &  & 0.00 & 0.00 & 0.00 \\
\hline
\hline
\end{tabular}
\end{center}
\end{table*}

In contrast, for the $^{30,32}$Si nuclei, the dependence on the parameter 
set and the treatment of pairing is much stronger. 
For instance, with the NLSH parameter set, the $^{30+\Lambda}$Si is slightly 
oblate and the deformation parameter is similar between 
$^{32}$Si and $^{32+\Lambda}$Si. Without the pairing correlation, the 
deformation is similar between $^{32}$Si and $^{32+\Lambda}$Si for both NL3 and NLSH. 
Apparently more careful investigations will be necessary for 
these nuclei before we can draw a definite conclusion on their deformation 
parameter. We summarize our results for $^{28,30,32}$Si and 
$^{28,30,32}$Si + $\Lambda$ in Table I. 

\begin{figure}[htb]
\begin{center}\leavevmode
\includegraphics[width=0.91\linewidth, clip]{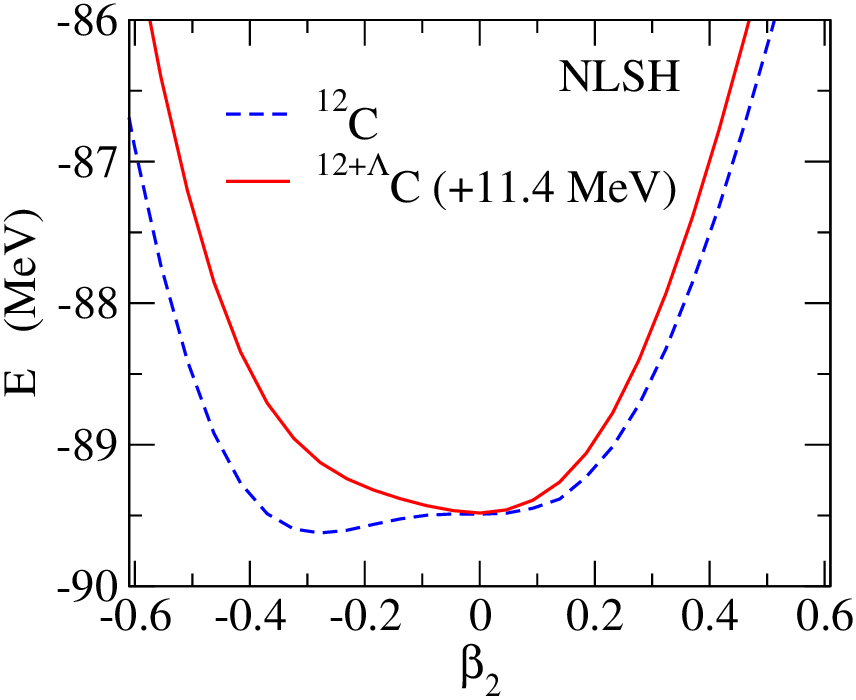}
\caption{(Color online) 
Same as Fig. 5, but 
for the $^{12}$C and $^{12+\Lambda}$C nuclei. 
}
\end{center}
\end{figure}

As another example which shows a large effect of $\Lambda$ particle 
on nuclear deformation, 
we next discuss the $^{12}$C nucleus. 
For this nucleus, the calculation with the NL3 parameter set did not 
converge, due to the instability of the scalar meson 
field \cite{R88,FBR89}, and we here show only the results with the NLSH set. 
Fig. 6 shows the potential energy surface obtained with the NLSH parameter 
set together with the constant gap approximation for the pairing correlation. 
The behaviour of energy surface of $^{12}$C is similar to that of $^{28}$Si 
shown in Figs. 4 and 5. That is, the energy surface 
has a shallow oblate minimum 
and a shoulder at the spherical configuration. 
For this nucleus, the energy difference between the oblate and the spherical 
configurations is as small as 0.13 MeV. By adding a $\Lambda$ particle, 
the oblate minimum disappears and the ground state becomes spherical. 
This is exactly the same effect of $\Lambda$ particle as that in the 
$^{28}$Si nucleus. For this light nucleus, the pairing correlation 
does not play an essential role, and we confirm that our conclusion 
remains the same even if we do not include the pairing correlation (see Table I). 

\section{Summary}

We have used the relativistic mean field (RMF) theory to investigate 
quadrupole deformation of $\Lambda$ hypernuclei. 
We have shown that, while an addition of $\Lambda$ particle does not 
influence much the shape of many nuclei, 
$^{12}$C and $^{28}$Si make important exceptions. 
That is, we have demonstrated that 
the $\Lambda$ particle makes the shape of these nuclei 
change from oblate to spherical. 
For the $^{28}$Si nucleus, this conclusion was 
achieved 
both with the NL3 and NLSH parameter 
sets of the RMF Lagrangian, although the calculation with NL3 
was not converged for the $^{12}$C nucleus 
due to the instability of sigma field.  
We have also 
confirmed that the conclusion is independent of the treatment of 
pairing correlation among the nucleons. 

An important next question will be how to observe experimentally 
the drastic structure change of the hypernuclei found in this paper. 
For this purpose, a measurement of the energy of the first 4$^+$ state, 
and thus a deviation from a rotational spectrum, 
will be extremely useful. On the other hand, the potential energy surface 
for the $^{13}_{~\Lambda}$C and $^{29}_{~\Lambda}$Si nuclei 
is somewhat soft and a large 
anharmonic effect of collective vibration might be expected. 
One may thus need to perform {\it e.g.,} a generator coordinate method (GCM) 
calculation \cite{BH08},  
on top of the mean field calculation presented in this paper, 
and calculate the excitation spectra before one can compare the theoretical 
results with experimental data. 

\begin{acknowledgments}
We thank H. Tamura, H. Sagawa, Nyein Wink Lwin, and Khin Nyan Linn for 
useful discussions. 
This work was supported by the Japanese
Ministry of Education, Culture, Sports, Science and Technology
by Grant-in-Aid for Scientific Research under
the program number 19740115.
\end{acknowledgments}


\begin{thebibliography}{99}

\bibitem{BHR03}M. Bender, P.H. Heenen, and P.-G. Reinhard, 
Rev. Mod. Phys. {\bf 75}, 121 (2003). 

\bibitem{SDNPD03}M.V. Stoitsov, J. Dobaczewski, W. Nazarewicz,
S. Pittel, and D.J. Dean, Phys. Rev. C{\bf 68}, 054312 (2003). 

\bibitem{BH08}M. Bender and P.H. Heenen, Phys. Rev. C, in press, 
and references therein. arXiv:0805.4383 [nucl-th].

\bibitem{R81}M. Rayet, Nucl. Phys. {\bf A367}, 381 (1981). 

\bibitem{YBZ88}Y. Yamamoto, H. Band\=o, and J. Zofka, Prog. Theo. Phys. {\bf 80}, 757. 

\bibitem{YB90}Y. Yamamoto and H. Band\=o, Prog. Theo. Phys. {\bf 83}, 254 (1990). 

\bibitem{LY97}D.E. Lanskoy and Y. Yamamoto, Phys. Rev. C{\bf 55}, 2330(1997). 

\bibitem{L98}D.E. Lanskoy, Phys. Rev. C{\bf 58}, 3351 (1998). 

\bibitem{CLS00}J. Cugnon, A. Lejeune, and H.-J. Schulze, Phys. Rev. C{\bf 62}, 064308 (2000). 

\bibitem{VPRS01}I. Vida\~na, A. Polls, A. Ramos, and H.-J. Schulze, 
Phys. Rev. C{\bf 64}, 044301 (2001). 

\bibitem{ZSSWZ07}X.-R. Zhou, H.-J. Schulze, H. Sagawa, C.-X. Wu, 
and E.-G. Zhao, Phys. Rev. C{\bf 76}, 034312 (2007). 

\bibitem{MZ89}J. Mares and J. Zofka, Z. Phys. A{\bf 333}, 209 (1989); {\it ibid}, A{\bf 345}, 47 (1993). 

\bibitem{RSM90}M. Rufa, J. Schaffner, J. Maruhn, H. St\"ocker, W. Greiner, and 
P.-G. Reinhard, Phys. Rev. C{\bf 42}, 2469 (1990). 

\bibitem{GVH93}N.K. Glendenning, D. Von-Eiff, M. Haft, H. Lenske, and M.K. Weigel, Phys. Rev. C{\bf 48}, 889 (1993). 

\bibitem{MJ94}J. Mares and B.K. Jennings, Phys. Rev. C{\bf 49}, 2472 (1994). 

\bibitem{ST94}Y. Sugahara and H. Toki, Prog. Theo. Phys. {\bf 92}, 803 (1994). 

\bibitem{VPLR98}D. Vretenar, W. P\"oschl, G.A. Lalazissis, and P. Ring, 
Phys. Rev. C{57}, R1060 (1998). 

\bibitem{LMZZ03}H.F. L\"u, J. Meng, S.Q. Zhang, and S.-G. Zhou, 
Eur. Phys. J. A{\bf 17}, 19 (2003). 

\bibitem{SYT06}H. Shen, F. Yang, and H. Toki, Prog. Theo. Phys. {\bf 115}, 325 (2006). 

\bibitem{HT06}O. Hashimoto and H. Tamura, Prog. Part. Nucl. Phys. 
{\bf 57}, 564 (2006). 

\bibitem{RGL90}
Y.K. Gambhir, P. Ring, and A. Thimet, Ann. of Phys. (N.Y.), 
{\bf 198}, 132 (1990). 

\bibitem{R96}P. Ring, Prog. Part. Nucl. Phys. {\bf 37}, 193 (1996). 

\bibitem{RGL97}P. Ring, Y.K. Gambhir, and G.A. Lalazissis, 
Comp. Phys. Comm. {\bf 105}, 77 (1997). 

\bibitem{RS80}
P. Ring and P. Schuck, {\it The Nuclear Many Body Problem}
(Springer-Verlag, New York, 1980).

\bibitem{HLY06}K. Hagino, N.W. Lwin, and M. Yamagami, 
Phys. Rev. C{\bf 74}, 017310 (2006). 

\bibitem{nl3}G.A. Lalazissis, D. Vretenar, and P.Ring, Phys. Rev
  C{\bf 55}, 540 (1997)

\bibitem{MN92}P. M\"oller and J.R. Nix, 
Nucl. Phys. {\bf A536}, 20 (1992). 

\bibitem{BG03}A. Bhagwat and Y.K. Gambhir, 
Phys. Rev. C{\bf 68}, 044301 (2003). 

\bibitem{LRR99}G.A. Lalazissis, S. Raman, and P. Ring, 
At. Data Nucl. Data Tables {\bf 71}, 1 (1999). 

\bibitem{FQKV73}H. Flocard, P. Quentin, A.K. Kerman, and D. Vautherin, 
Nucl. Phys. {\bf A203}, 433 (1973). 

\bibitem{SNR93}M.M. Sharma, M.A. Nagarajan, and P. Ring, 
Phys. Lett. B{\bf 312}, 377 (1993). 


\bibitem{RNS78}I. Ragnarsson, S.G. Nilsson, and R.K. Sheline, 
Phys. Rep. {\bf 45}, 1 (1978). 

\bibitem{R88}P.-G. Reinhard, Z. Phys. {\bf A329}, 257 (1988). 

\bibitem{FBR89}J. Fink, V. Blum, P.-G. Reinhard, J.A. Maruhn, 
and W. Greiner, Phys. Lett. B{\bf 218}, 277 (1989). 

\end{thebibliography}
\end{document}